\documentclass{article}
\usepackage[utf8]{inputenc}
\usepackage[english]{babel}
\usepackage{csquotes}
\usepackage[margin=1in]{geometry}

\font\tfont=cmss12 at 32pt
\font\hfont=cmss12 at 20pt
\begin{document}
  \title{{\tfont A Survey of Advances in Botnet Technologies}}
  \date{{\hfont 2017-01-15}}
  \author{{\hfont Nathan Goodman}}
  \pagenumbering{gobble}
  \maketitle
  \pagenumbering{arabic}
  \newpage

\section{Abstract}

Botnets have come a long way since their inception a few decades ago. Originally toy programs written by network hobbyists, modern-day botnets can be used by cyber criminals to steal billions of dollars from users, corporations, and governments. This paper will look at cutting-edge botnet features and detection strategies from over a dozen research papers, supplemented by a few additional sources. With this data, I will then hypothesize what the future of botnets might hold.

\section{Introduction}

A botnet is a network of computers that are controlled by a single machine called a botmaster. A user typically becomes infected when they open a malicious email attachment, visit a compromised website, or unsuspectingly download the bot onto their computer. Once infected, the botmaster will gain access to the victim’s computer, unpronounced to the victim. \cite{kim_jeong_kim_so_2010, tanwar_goar_2014}

Being in control of a botnet gives the botmaster two critical resources: CPU power and IP addresses. Even if a bot only makes use of 5\% of a machine’s CPU resources, aggregated amongst hundreds of thousands of machines, this modest amount can give the botmaster the power equivalent of a modern-day supercomputer. But unlike a supercomputer which would execute traffic on a single IP address, a botmaster can make it look as if traffic were coming from thousands of distinct sources, that being its victims. \cite{silva_silva_pinto_salles_2013}

This makes botnets an extremely effective tool for performing malicious network operations. Some common examples include distributed denial of service (DDoS) attacks, spam campaigns, and click fraud (clicking on pay-per-click ads in order to increase profits). Although the raw CPU power to perform these malicious actions could be obtained on a single powerful (albeit expensive) machine, the diversity of IP addresses is necessary for the criminals to avoid detection. A network administrator can easily block a single IP address, but cannot easily block thousands of IP addresses, especially when they all appear to be legitimate. \cite{kim_jeong_kim_so_2010, rodriguez_gomez_macia_fernandez_garcia_teodoro_2013}

Additionally, botnets can be used as a back-door to steal personal information from their host machines. They can either trick users from entering their information, or even capture information using tools such as keyloggers and screenshots. And because the botnet already has system privileges victim’s computer, they can easily send the data back to the botmaster. \cite{rodriguez_gomez_macia_fernandez_garcia_teodoro_2013}

As the number of internet-connected devices grows exponentially, botnets have grown to become one of the greatest threats to global security. According to industry estimates, botnets have caused over \$110 billion in losses globally \cite{demarest_2014}. The FBI considers botnets to be one of the greatest threats to the cyber security of the US government and American industries \cite{demarest_2014}. And they are also becoming increasingly prevalent. Although there are no perfect metrics for measuring the number of infected machines at any given time, by most estimates around 1 in 5 computers worldwide is part of a botnet \cite{silva_silva_pinto_salles_2013}. Depending on the sophistication of the botnet, anti-virus software may or may not be able to detect it. And once a bot is there, removing it can be incredibly difficult. \cite{kim_jeong_kim_so_2010, stone_gross_cova_cavallaro_gilbert_szydlowski_kemmerer_kruegel_vigna_2009}

Botnet technology has advanced rapidly in the past decade. The first botnets were simple programs that executed on a machine’s startup that used IRC (internet relay chat) to communicate with the C\&C (command and control server). These types of botnets were generally easy to detect, simple to remove, and limited in the amount of damage they could cause. In contrast, today’s botnets use protocols such as peer-to-peer (P2P) or HTTP, which are far more robust, difficult to block, and mimic legitimate traffic. They tend to bury themselves into a computer’s file system, making them hard to detect and difficult to remove. And they have become increasingly powerful too, as botmasters find new ways to exploit them to carry out attacks. \cite{sood_enbody_bansal_2013}

The motives of botnets have also changed considerably. In the early days, botnets were usually constructed by network hobbyists and rarely caused serious harm. But today, most botnets are constructed primarily for financial gain. Understanding this is imperative to understanding the future direction of botnet technology. \cite{stone_gross_cova_cavallaro_gilbert_szydlowski_kemmerer_kruegel_vigna_2009}

The question going forward is not whether botnet technology will continue to advance, but rather what it will look like. This paper will present a number of studies which focus on the future of botnet technologies. These can be grouped into one of three categories: new designs, new targets, and new attack types. Afterward, this paper present studies which describe methodologies of detecting botnets, helping to reduce the damage they cause.

\section{Botnet Studies}

\subsection{Design}

Early botnets all used a centralized model, where all of the bots in the botnet report to a single server. This strategy is simple to construct and achieves a high latency. Unfortunately, it also leaves the botnet extremely vulnerable. Having a single point of failure means that taking down the botnet is as simple tracking down the IP address of the C\&C server, and blacklisting it.

To solve this problem, botnet designers began creating botnets using P2P technology. The first such botnet to do this was Nugache, back in 2006. When a user gets infected, the program starts and automatically connects with other victims. When the botmaster decides to use the botnet, they relay commands to a few infected machines, which then relay the commands another set of bots, and so on, until the command eventually propagates across the entire network. \cite{bailey_cooke_jahanian_xu_karir_2009, silva_silva_pinto_salles_2013}

The primary advantage of this system is that the vast majority of bots don’t communicate directly with the commanding machine. Therefore, its IP address cannot easily be identified, and it cannot easily be shut down. However, the usage of the P2P network adds considerable latency since it takes time for a command to propagate, especially with larger networks. But as the cyber security community invests more resources into taking down botnets, most future botnets will likely use some kind of P2P structure to avoid being compromised. \cite{silva_silva_pinto_salles_2013}

In one study \cite{sanatinia_noubir_2015}, researchers developed a concept for an even stronger botnet design known as OnionBots. An OnionBotnet emulates the basic concept of a P2P botnet, but makes use of Tor, an open source anonymity platform, to hide its location. Instead of bots storing neighboring IPs like traditional P2P botnets, bots in an OnionBotnet store .onion addresses of other botnets. Due to Tor’s anonymity structure, these .onion addresses cannot easily be traced back to their source. And since most current P2P botnet mitigation techniques require IP addresses of a vast number of bots, this design renders these techniques invalid.

Although most new botnets employ P2P at some level of their design, others are finding alternative means to overcome the “single point of failure” issue. In the case of Torpig, the designers implemented a technique called domain flux, which associates multiple domain names with the C\&C server. \cite{stone_gross_cova_cavallaro_gilbert_szydlowski_kemmerer_kruegel_vigna_2009, silva_silva_pinto_salles_2013}

To understand domain flux, it is necessary to first understand IP flux. Traditionally, the C\&C server of “central command” botnets contained only a single IP address. Should the IP address of a the C\&C be discovered, it could easily be blacklisted, thereby disabling the bot. In order to get around this, cyber criminals would purchase a set of IP addresses. Should one IP address be compromised, the bots could simply be programmed with backup IPs that will also connect them to the C\&C, thus implementing IP flux.

However, most botnets use DNS to resolve their IP addresses, and wiring up multiple IP addresses to a single DNS name still constitutes a single point of failure. Instead of blocking the botnet at the IP level, the botnet can now be blocked at the DNS level. While this can be more difficult since it requires collaboration with DNS companies, most of them will block a DNS entry if presented with proof that it’s being used with malicious intent. \cite{stone_gross_cova_cavallaro_gilbert_szydlowski_kemmerer_kruegel_vigna_2009, silva_silva_pinto_salles_2013}

Therefore, some modern botnets have implemented domain flux. Domain flux is a system by which a botnet uses an algorithm to determine the set of possible domains. If a bot cannot find the C\&C at its previous domain, it will search over the possible DNS space until it finds it. Using this method, should one DNS entry be shut down, the botmaster can simply alter the DNS associated with the C\&C server, and the botnet will continue to function as normal. Should the botmaster be running low on domains, they can simply update the algorithm which determines DNS space, dynamically expanding its possible resolutions. In this way, an actively maintained botnet can usually keep up with attempts to take it down, while still maintaining high-latency. \cite{stone_gross_cova_cavallaro_gilbert_szydlowski_kemmerer_kruegel_vigna_2009, silva_silva_pinto_salles_2013}

However, domain flux can also leave the botnet vulnerable. If an attacker can reverse engineer the domain generation algorithm, they can potentially register one of the possible domains and take control of the botnet. In theory, botmasters could eliminate this concern by purchasing all of the possible domains, but the financial cost of this would be prohibitive to most. In one study \cite{stone_gross_cova_cavallaro_gilbert_szydlowski_kemmerer_kruegel_vigna_2009}, researchers exploited this weakness and managed to take control of the Torpig botnet for 10 days.

As mentioned earlier, the biggest factor involved in botnet design today is money. As botnets have become a multi-million dollar business, every aspect of the botnet process has become commercialized. This has led to the phenomenon of BaaS (Botnet as a Service), modeled after the more well-known SaaS (Software as a Service). \cite{chang_wang_mohaisen_chen_2014}

In the past, the botnet process was usually localized. What this means is if advisory A was running a spam campaign, then more likely than not, it was also advisory A that harvested the emails, constructed the botnet, designed the actual spam email, and executed the attack. But today, all of those people could be different. A professional spammer could log onto the darknet and buy an email list, rent out a botnet, and execute the attack themselves. The significance of this is that it makes botnets far more accessible – a cyber criminal no longer needs the technical skills to make a botnet in order to use one. This opens up cyber criminal activity to less tech-savvy individuals. \cite{chang_wang_mohaisen_chen_2014}

Given the popularity of BaaS products, many botnets today as designed as such. A botnet designer may have no purpose for the botnet itself, but simply wish to profit by renting it cyber criminals. These botnets are usually more flexible. They can be easily handed over to a purchaser, and can often be easily customized. A good example of this type of botnet is SkyEye. Once it infects a computer, it installs a BDK (bot development kit) that the purchaser can use to extend the functionality of the botnet. \cite{sood_enbody_bansal_2013}

But BaaS isn’t the only way botnet designers are making money. A new phenomenon known as DIY botnets has resulted in botnets becoming increasingly accessible to the average cyber criminal. Botnet designers will sell their software as user-friendly packages that anybody can set up. This has resulted in what researchers refer to a “mini-bots”, or many small identical botnets produced from the same code, under the control of different masters. Although not usually as powerful as traditional botnets, DIY botnets are cheaper for the criminal than BaaS and still require less technical skills than creating a botnet from scratch. Zbot is one of the most successful examples of this phenomenon, with over 500 mini-botnets at its peak in the late 2000s. \cite{al_bataineh_white_2011}

\subsection{Attacks}

Phishing is an attack in which cyber criminals trick users into giving them their personal information by masquerading as legitimate sources. Phishing attacks have been around for years and botnets have long been involved in the process. However, new advancements in phishing techniques are making it both increasingly dangerous and difficult to detect.

The Torpig botnet is an example of a botnet which can generate devastating phishing attacks. Unlike other botnets which primarily use their host to carry out other attacks, one of Torpig’s primary functionalities is stealing personal information from its victims. Once installed on the victim’s computer, it downloads plugins which automatically steal data from a number of popular applications. This in and of itself can be devastating if the victim uses the machine for tasks which require exposing important personal information, such and banking or taxes. \cite{stone_gross_cova_cavallaro_gilbert_szydlowski_kemmerer_kruegel_vigna_2009}

But Torpig then uses this information to phish for additional information. When a user navigates to a legitimate website, such as their online banking account, Torpig can redirect them to a malicious web page which asks them to confirm their personal information. These pages often show the same URL as the real one, and often have the exact same look and feel as the legitimate page. And because Torpig already has information from the user, they can use this to make the page feel even more real. Once the user enters their information and submits it, they’re usually forwarded to a legitimate page, making them completely unaware that an attack just occured. Unlike traditional phishing attacks, which often fail to trick educated users, these types of integrated phishing attacks will fool almost anyone. \cite{stone_gross_cova_cavallaro_gilbert_szydlowski_kemmerer_kruegel_vigna_2009, sood_enbody_bansal_2013}

Email harvesting is the process of obtaining lists of legitimate email addresses through HTTP web crawlers, usually for spam campaigns. The email addresses can either be harvested directly by the spammer, or sold on the darknet to cyber criminals executing spamming attacks. Although it is possible to guess addresses, a campaign with legitimate addresses will be far more successful and far more profitable. And while many think that spam is no longer a big deal since spam filters have improved, a successful spam campaign can still generate over half a million dollars in revenue. As such, spam isn’t going away anytime soon. \cite{stringhini_hohlfeld_kruegel_vigna_2014}

Increasingly, botnets are being used to carry out email harvesting. An analysis in \cite{stringhini_hohlfeld_kruegel_vigna_2014} determined that a single email harvester contained 56 IP addresses within a 24 hour period, all of which were spread around Germany. While this could technically signify any sort of distributed structure, it is very likely to have been a botnet.

What’s unique about this scenario is that a botnet is not currently required to harvest email addresses. While it is possible that a site could block an IP address for suspicious traffic, in practice most websites are not monitored closely enough to do this. The same study also found several email harvesters that were operating on a single IP address, and many of them were quite successful. But as webmasters become increasingly savvy about stopping web crawlers, we could see botnets increasingly employed as email harvesters.

Cryptocurrencies such as BitCoin have exploded in value over the last few years. And as mining, or the process of obtaining new currency, becomes increasingly profitable, the incentive to mine only grows. This has resulted in botmasters using their network to mine bitcoin and other cryptocurrencies. Infected machines can be installed with the mining software, which is free and open source. Once installed, botnets start up the mining software in the background while the infected computer is on, a process which is invisible to the user. \cite{huang_mccoy_dharmdasani_meiklejohn_dave_grier_savage_weaver_snoeren_levchenko_et}

There are two different ways a botnet can be exploited for mining. The first is by joining a pool, or a collective mining group which shares its collective loot amongst its members. In order to avoid detection, botmasters can easily proxy their bots into a single IP, making them seem like a single, powerful computer. Most pools will split the earnings up by contributed CPU power, so a botnet can earn an increasing amount of bitcoin as its network grows in size. Alternatively, a botmaster can construct a dark pool, or a pool owned by the botmaster. These types of botnets typically have hundreds of thousands of members, having the necessary CPU power to form their own pool. Regardless of the pool type, any earnings the botnet obtains can be exchanged for standard currencies in an anonymous fashion.  \cite{huang_mccoy_dharmdasani_meiklejohn_dave_grier_savage_weaver_snoeren_levchenko_et}

\subsection{Target}

Within the span of a few decades, smartphones have gone from science fiction to ubiquity. We rely on smartphones so much in our daily lives, that many of us forget that this technology is still very new. As such, a number of security flaws in mobile OSes remain open. Combined with OSMs (Online Social Media), which collectively have billions or users, and the result is a combination that could turn into a powerful botnet.

SoCellBot is a proposed botnet that specifically targets mobile OSes. It propagates via OSMs, sending users a benign-looking link to click on, potentially masquerading as a user’s close friend. This would allow the malware to spread quickly, potentially to hundreds of millions of devices globally. Once infected, it can exploit vulnerabilities in the mobile OS to install onto the phone, thereby infecting the device. As mobile devices become increasingly powerful, the strength of such a botnet could potentially rival a traditional botnet. \cite{faghani_nguyen_2012}

Additionally, OSM botnets can be used to obtain personal information. On most social media platforms, a user’s personal information is hidden to the general population. Yet friends, and friends of friends, can view certain pieces of personal information such as name, sex, email, and phone number. While most users don’t mind connections seeing this information, a malicious user can exploit this information to track, spam, or try to obtain more valuable information from a user. Once the botnet connects with a user, it can relay all of its data back to its botmaster. And even if a user’s connection connects with a bot, and that user does not connect, their information could still be stolen depending on their privacy settings. \cite{boshmaf_muslukhov_beznosov_ripeanu_2013, faghani_nguyen_2012}

But smartphone are not the only new “smart” devices that are vulnerable. According to Proofpoint, a cyber security company, they’ve discovered the very first botnet that attacked a smart refrigerator \cite{your_fridge_is_full_of_spam}. It appears as though the malware wasn’t written specifically for the device, but rather was written for a generic Linux system which happened to infect the device. Although this is just an isolated case, it shines light on the possibility of such attacks in the future. As an increasing number of our everyday devices become connected to the internet, it’s likely that cyber criminals will attempt to exploit them for malicious purposes.

\subsection{Detection}

An all-out arms race has taken off between the cyber criminals creating new botnet technologies, and the cyber security community which is trying to destroy them. Given enough time and resources, a botnet can often be brought down. But taking it down requires studying the botnet, finding security holes, and executing the attack (or counter-attack). And in order to accomplish this, the first step is simply detection the botnet. While it may sound simple to detect a network of hundreds of thousands of bots, advances in botnet technology make this task increasingly more difficult.

Historically, the most common way to detect a botnet was via the signature method. Once a single copy is found, a botnet’s binary signature can be stored. This can then be used to screen additional systems for the malware by simply checking the bot’s binary signature. But while signature detection can be an effective tool, it also has some enormous drawbacks. For one, every time a botnet is updated, its signature changes. Since most modern botnets have built-in updating mechanisms, a botmaster can simply update the malware to change its signature. Additionally, many modern botnets will have multiple versions which are identical in functionality, but differ in signature. This renders signature detection significantly more challenging. \cite{zhao_traore_sayed_lu_saad_ghorbani_garant_2013, landecki}

A newer mechanism for detecting botnets is being studied called network flow. It is based on the principle that botnet traffic is distinct from legitimate web traffic. For example, botnet traffic is typically quite uniform amongst its bots, often occurs at specific intervals, and the number of packets exchanged between the C\&C and the bot is roughly uniform each time. Traditional web traffic is obviously much more random, meaning that these patterns can signify a high likelihood of a botnet. Most importantly, network flow analyses are immune to encryption, which is essential given the ubiquity of encryption today. Machine learning can be used to fine-tune detection algorithms and detect changes in botnet network patterns. \cite{li_liu_cui_2014, lee_chou_chen_2015}

Another promising detection technique is group network analysis. By the very definition of a botnet, a large number of machines must be performing the same kind of network operations aronud the same time. By aggregating this into a giant matrix, botnets can be uncovered by determining uniformity across the network. In \cite{zhao_traore_sayed_lu_saad_ghorbani_garant_2013}, researchers developed BotGAD, a group network analysis tool that was able to detect 20 unknown botnets in a controlled environment on a college campus. 

Unfortunately, both of these techniques have some serious flaws. Stealthy botnets can be designed to be more difficult to detect via network flow, by introducing randomization into their communication algorithm. While this does constrict the botnet, usually reducing its efficiency, larger botnets can easily afford the cost. Group network analysis is more difficult for a botnet to evade, but it becomes impractical as the network sample size increases. So while it remains a promising technique, it is difficult to apply to real-world networks. And unlike signature checking, these techniques often result in both false positives and false negatives. Legitimate software that follow botnet-like patterns, such as update managers, can be easily mistaken for botnets. \cite{zhao_traore_sayed_lu_saad_ghorbani_garant_2013}

Rather than relying on just a single method, novelty botnet detection systems being proposed take into account a number of detection systems. An evaluation scheme, described in \cite{zhao_traore_sayed_lu_saad_ghorbani_garant_2013}, proposes using signature analysis to catch established threats, and network flow analysis to identify new or lesser known threats. Additionally, the system would whitelist legitimate traffic that could often be construed as a botnet. Using these methods, they hypothesize that such a system could achieve a detection rate of over 90\% (assuming modern botnet technology) and a false positive rate of under 5\%. \cite{li_liu_cui_2014}

\section{Conclusion}

Botnets are one of the greatest cyber threats of our time, and their prominence is only increasing. New design features, attack methods, and targets will likely make the next generation of botnets as dangerous as ever. On the other hand, detection and mitigation strategies are also on the rise. We can only wait and see which side will come out ahead.

Future botnets will probably leave the central command structure behind. Faced with the threat of active mitigation, these botnets simply will not hold up. Even with advances such as IP and domain flux, central command botnets are vulnerable and expensive to maintain.  Therefore, future botnets will probably use P2P technology in at least some respect. So-called OnionBots, or bots with built-in Tor anonymity, could conceivably become the standard for future botnets. Additional advances will likely focus on improving network latency, which will increase the power and therefore profitability of the botnet. And as cyber detectives increasingly try to detect botnets, cyber criminals will probably invest more in detection evading techniques.

We can also expect botnets to encompass a more modular design. This would allow them to be easily sold and distributed to other cyber criminals, making botnets increasingly more accessible. We will probably also see an increasing separation between the roles of a botnet operation. For example, a spam campaign may involve one person who harvests the emails, another who designs the botnet, another who designs the phishing attack, and a person who pays for all those services, then executes the botnet. While today it is possible to obtain botnets deep into the darknet, it could someday be possible to google “botnet”, download an executable file, and start infecting other machines. Such a situation would undoubtedly increase the number of botnets out there, and present a security challenge unlike any in our time.

The next generation of botnets will also likely execute a far wider range of attacks than present day botnets. While traditional operations such as DDoS, click fraud, and spam will continue to be used, newer botnets will also perform targeting phishing attacks, mine bitcoin, and harvest emails. As new software becomes increasingly integrated into our lives, I think we’ll see new types of botnet attacks that we can’t even imagine today. And as botnets becoming increasingly versatile, their value will only increase as their potential financial gains increase.

Despite the modern-day ubiquity of mobile devices, botnet designers have been slow to tap into this market. My hypothesis is that in the past, the CPU power of these devices was simply marked off as insignificant. But the smartphones of today have wildly faster processors than just a few years ago, and the number of smartphones out there has grown exponentially. In all likelihood, it’s simply a matter of time before this hole becomes properly exploited. And as IoT (internet of things) technology takes off, other devices such as televisions, refrigerators, and light bulbs will all become vulnerable to botnets.

On the bright side, researchers are becoming increasingly able to detect botnets. And once a network is detected, it becomes significantly easier to destroy. Network flow and group analysis detection methods, especially if combined with other detection mechanisms, look like promising tools for detecting the next generation of botnets. However, given the knowledge of these methods, botnet designers can construct protocols which can greatly reduce the likelihood of detection. We desperately need better detection techniques if we are going to win the arms race against the cyber criminals. Otherwise, we could see a world with a lot more botnets, and a lot less security.

\bibliographystyle{abbrv}
\bibliography{bibliography}

\begin{thebibliography}{10}
\bibitem{your_fridge_is_full_of_spam}
Your fridge is full of spam: Proof of an iot-driven attack.
\end{thebibliography}

\end{document}